\documentclass[11pt,a4paper]{article}

\usepackage{epsfig,a4}

\def\color[#1]#2{}

\def\anglefitA{0.22}
\def\anglefitB{6.473}
\def\parfitRMS{2.2}
\def\parfitchi2{4.9}
\def\anglefitposRMS{0.28}
\def\anglefitposchi2{0.08}

\def\dmath#1#2{
$$\lineskiplimit=1000pt \advance\lineskip by #1\jot 
\mathsurround=0pt \tabskip=0pt plus 1000pt
\everycr{\noalign{\penalty\interdisplaylinepenalty}}
\halign to \displaywidth{
\hfil$\displaystyle{##}$\tabskip=0pt&%
\hfil $\displaystyle{{}##{}}$\hfil &%
$\displaystyle{##}$\hfil \tabskip=0pt plus 1000pt minus 1000pt&%
\refstepcounter{equation}\label{##}\llap{(\theequation)}\tabskip=0pt\cr
\noalign{\ifdim \prevdepth>-1000pt \vskip -#1\jot\fi}
#2\crcr}$$}

\def\crl#1{\crcr\noalign{\unpenalty\penalty 10000
\nointerlineskip \vbox to 0pt {
\dimen0=\lineskip \vskip \dimen0 minus 1000pt \hbox to \displaywidth{%
\hfil \refstepcounter{equation}\label{#1}(\theequation)}
\vskip 0pt minus 1000pt} \penalty 10000}}

\begin{document}

\begin{center}
  {\Large \bf Numerical investigation of flow separation in the lee side of
    transverse dunes}

\vskip 8mm

{\large Volker Schatz$^1$ and Hans J. Herrmann$^{1,2}$}\\
{\em $^1$ Institute for Computational Physics, Stuttgart University, Pfaffenwaldring 27, 70569 Stuttgart, Germany \\
$^2$ Departamento de F\'\i sica, Universidade Federal do Cear\'a, Fortaleza, Brasil}

\end{center}

\begin{abstract}
  We investigate flow separation in the air flow over transverse sand dunes.
  CFD simulations of the air flow over differently shaped dunes are performed.
  The length of the recirculation region after the brink of the dune is found
  to depend strongly on the shape of the dune.  A phenomenological expression
  for the separation length is presented.  Suitably non-dimensionalised, it
  depends linearly on the angle of the dune at the slip face brink.  We
  propose a fit for the shape of the separating streamline, which is well
  approximated by an ellipse.
\end{abstract}


\section{Introduction}

Dunes are naturally occurring, beautifully shaped sand deposits.  Since the
middle of the previous century, they have attracted the attention of scientists
who have been seeking to model them and understand the processes leading to
their formation.  From the point of view of the physicist, sand dunes
constitute a variable boundary problem: The air flow is determined by the shape
of the dune and in turn influences the dune shape by transporting sand grains.
Therefore the air flow over dunes is of great importance for understanding dune
formation and evolution.  Consequently, this topic has aroused a great deal of
interest since the days of Bagnold~\cite{Bagnold41,Bagnold51} and led to a
significant number of publications
\cite{Benjamin59,Sutherland67,Brown79,Haff83,McLean86,Rubin87,Wood95,NelsonSmith89,Weng91,Ayotte04}.

Since the start of scientific interest in dunes, there has been some work on
the topic of flow separation in the lee of dunes, both
theoretical~\cite{NelsonSmith89,Parsons04} and experimental (e.\ g.\
\cite{Engel81,SweetKocurek90,Rasmussen96}).  However, due to the difficult
nature of the problem, these papers have only tackled part of the problem.  In
several publications, transverse dunes have been modelled as triangular
structures~\cite{Engel81,Parsons04,Parsons04a}.  Field measurements of air flow
over dunes, on the other hand, tend to lack measurements of the dune
profile~\cite{SweetKocurek90,FrankKocurek96}.

A recent field measurement~\cite{Parteli04} suggests that the shape of
transverse dunes has significant influence on the length of the recirculation
region.  Since the sand transport in the recirculation region in the lee of a
dune is negligible, the foot of the following dune shape is located at or
downwind of the flow reattachment point (if one assumes the dune shapes to be
stable).  Therefore the distance of closely spaced dunes is a limiting measure
of the length of the recirculation region.  In reference~\cite{Parteli04} the
separation length after different dunes was determined in this way.

In this paper we will present results for widely spaced or isolated transverse
dunes.  This is to some extent an idealisation.  However, we think it is a
useful idealisation: We want to concentrate on the effect of the dune shape
of a single dune, and taking into account the presence and shape of
neighbouring dunes would introduce additional parameters.  In
Section~\ref{sec:close} we discuss the effect of considering transverse dunes
which are part of a dune field.

This text is organised as follows: In the following Section~\ref{sec:method}
the models and parameters of our CFD simulations are described.  The geometry
of the dune shapes we modelled is also presented there.
Section~\ref{sec:seplen} presents our results for the length of flow separation
and the phenomenological formula we found.  In Section~\ref{sec:sepline} the
shape of the separating streamline extracted from the simulation is modelled
mathematically.  In Section~\ref{sec:close} we briefly discuss the situation of
a transverse dune in a field of closely spaced dunes.
Section~\ref{sec:discuss} compares our results with previous work.  The last
section presents a summary.

\begin{figure}
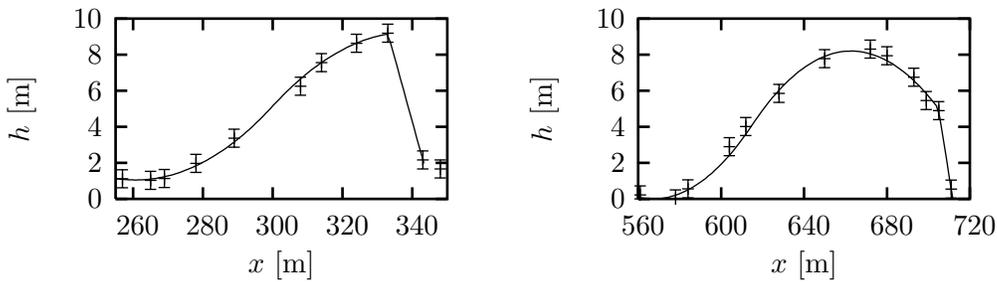

\hbox to \textwidth{
\hfill
\input{lencomp4.ptex}
\hfill
\input{lencomp7.ptex}
\hfill}
\caption{Realistic profiles of transverse dunes can be described approximately
         by two circle segments.  Data from \protect{\cite{Parteli04}}.}
\label{fig:lencomp}
\end{figure}

\section{Method}
\label{sec:method}

Our simulations were performed with the computational fluid dynamics software
FLUENT \cite{FLUENT}.  This software simulates the Reynolds-averaged
Navier-Stokes equations complemented by a turbulence model.  The simulations
were two-dimensional.  This implies translationally invariant dune shapes, i.\
e.\ perfectly straight transverse dunes, and a wind direction perpendicular to
the dunes.  The simulation grid was square.  We refined it near the ground to
allow a modelling of near-wall flow as accurate as possible using wall
functions.  Second-order discretisation schemes were used for all quantities
for which this was possible.

Besides the Reynolds-averaged Navier-Stokes equations, an additional set of
equations called the turbulence closure is required to determine a solution.
We use the $k$-$\epsilon$ model with renormalisation group (RNG) extensions.
This variant of the $k$-$\epsilon$ model was found to yield the most accurate
results in flow separation situations
\cite{Lien94,Bradbrook98,WalkerNickling02}.

The cross sections of the dune shapes were constructed from two circle
segments, a concave one modelling the foot of the dune and a convex one for the
crest.  This shape was chosen for reasons of convenience --- the program we
used to create the geometry supports circle segments.  But as can be seen from
Figure~\ref{fig:lencomp}, our shape provides a reasonable fit for real dunes.
The figure displays the dunes number 4 and 7 from Reference~\cite{Parteli04}.
Given the great variety of shapes found in nature, the measured shape may not
be universal.  But our geometric construction reflects the fact that the dune
profile is curved upward at its foot and downward at its crest and therefore
constitutes an improvement over the triangular shapes used previously.

\begin{figure}
\hbox to \textwidth{\hss\input{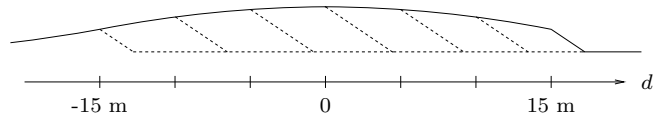}\hss}
\caption{The seven different dune shapes investigated.  The scale displays the
  brink position $d$.  The crest height of the dunes with positive brink
  position is 3 metres; the height of those with negative brink position equals
  the brink height, which is the smaller the more negative $d$ is.}
\label{fig:shape}
\end{figure}

To obtain different shapes, the position of the slip face was varied from the
start to the end of the convex part, see Figure~\ref{fig:shape}.  Note that
this has the consequence that not all the dunes have the same height.  The
heights and other geometrical data are given in Table~\ref{tab:geometry}.

The simulation results for the length of flow separation, our main quantity of
interest, was found to depend slightly on the spacing of the simulation grid.
To account for this small grid dependence, we performed the simulation of the
flow over each dune with three different grid sizes and extrapolated the
separation lengths to the continuum.  The average grid spacings were 10, 7 and
5~cm, respectively.

\begin{figure}
\input{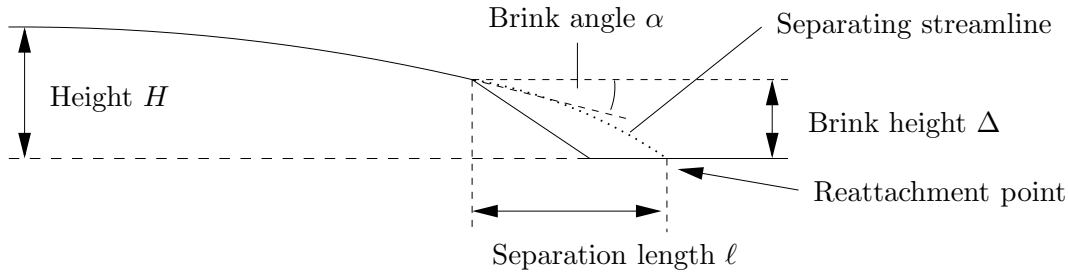}
\caption{The geometric variables characterising the dune shapes.  The brink angle is positive for dunes with a sharp brink and negative for round dunes as the one shown in this figure.}
\label{fig:geometry}
\end{figure}

\begin{table}
{\offinterlineskip
\def\tfil{\hskip 5pt plus 1fil \relax}
\halign{\vrule height2.5ex depth1ex width 0.7pt \tfil#\qquad\qquad&\vrule\qquad \tfil#&#\tfil 
&\vrule\qquad\quad  \tfil#&#\tfil &\vrule\qquad\quad\tfil#&#\tfil \vrule width 0.7pt\cr
\noalign{\hrule height 0.7pt}
\omit\vrule height3ex depth 1.5ex width 0.7pt\tfil Brink position $d$ [m] \tfil\ 
&\omit\span\omit\vrule\tfil Height $H$ [m] \tfil 
&\omit\span\omit\vrule\tfil Brink height $\Delta$ [m] \tfil 
&\omit\span\omit\vrule\tfil Brink angle $\alpha$ [$^\circ$] \tfil \vrule width 0.7pt \cr
\noalign{\hrule height 0.7pt}
$-$15   & 1&.5   & 1&.5   &  11&.4   \cr \noalign{\hrule}
$-$10   & 2&.337 & 2&.337 &   7&.6   \cr \noalign{\hrule}
$-$5    & 2&.835 & 2&.835 &   3&.78  \cr \noalign{\hrule}
0       & 3&     & 3&     &   0&     \cr \noalign{\hrule}
5       & 3&     & 2&.835 & $-$3&.78  \cr \noalign{\hrule}
10      & 3&     & 2&.337 & $-$7&.6   \cr \noalign{\hrule}
15      & 3&     & 1&.5   & $-$11&.4   \cr \noalign{\hrule height 0.7pt}
}}
\vskip 3pt
\caption{Geometric variables of the simulated dunes.  See
Figure~\protect{\ref{fig:geometry}} for a definition of the geometric
variables.  The brink angle is defined to be positive if the upwind slope is
positive at the brink.}
\label{tab:geometry}
\end{table}

\begin{figure}[h]
\hbox to \textwidth{\hss\input{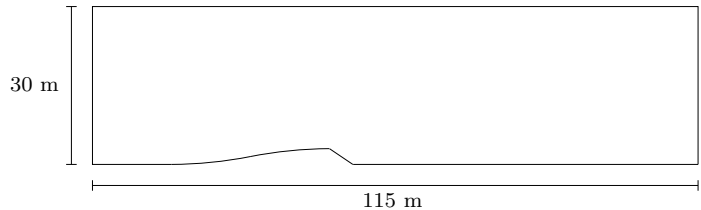}\hss}
\caption{The simulation region around the dune.}
\label{fig:simregion}
\end{figure}

The region around the dune in which the flow was simulated was chosen large
enough so that the boundaries did not influence the results.  This was verified
by performing simulations with larger simulation areas for some dune shapes and
comparing the results.  The simulation region extends 45\,m to the left and
70\,m to the right from brink position 0 (see Figure~\ref{fig:simregion}).  The
height of the simulated region was chosen to be 30\,m for all dunes except the
one with the most negative brink position which had the smallest height, where
20\,m was found to be sufficient.

The velocity profile at the influx boundary of the simulation region was set to
the logarithmic profile which forms in flow over a plane in neutral atmospheric
conditions:
\dmath1{
v(z) &=& \frac{u_*}{\kappa} \ln \frac z{z_0}\,,
&eq:logprof}
where $\kappa\approx0.4$ is the von K\'arm\'an constant.  The shear velocity
was chosen to be $u_*=0.4$ m/s.  The size of the roughness elements on the
ground, i.\ e.\ the sand grains, was chosen as $250\,\mu$m.  The roughness
length is 1/30 of the grain size, $z_0\approx 8.33\,\mu$m
\cite{Bagnold41,Wright97}.

\section{The flow separation length}
\label{sec:seplen}

The length of flow separation, our quantity of interest, was measured from the
slip face brink, where the flow separates, to the flow reattachment point (see
Figure~\ref{fig:geometry}), defined to be the position at which the velocity
near the ground changes direction from against the flow to in flow direction.
The separation lengths determined from simulations with different grid spacings
were extrapolated to the continuum with the standard linear regression
formulas.  To non-dimensionalise the separation length $\ell$, it was divided
by the height of the slip face.  Table~\ref{tab:seplen} shows the results for
all simulated dunes.

\begin{table}
\hbox to \textwidth{\hss\vbox{\offinterlineskip
\def\tfil{\hskip 5pt plus 1fil \relax}
\def\p{\,$\pm$\,}
\halign{\vrule height2.5ex depth1ex width 0.7pt \quad \tfil#&#\tfil 
&\vrule\quad \tfil#&#\tfil 
&\vrule\quad \tfil#&#\ \ \tfil \vrule width 0.7pt\cr
\noalign{\hrule height 0.7pt}
\omit\span\omit\vrule height2.5ex depth 1ex width 0.7pt\tfil 
Brink angle $\alpha$ [$^\circ$] \tfil\ 
&\omit\span\omit\vrule\tfil Separation length $\ell$ [m]  \tfil 
&\omit\span\omit\vrule\tfil  $\ell/\Delta$ \tfil \vrule width 0.7pt \cr
\noalign{\hrule height 0.7pt}
 11&.4    & 13&.22\p 0.5   &  8&.81\p 0.33 \cr \noalign{\hrule}
  7&.6    & 19&.16\p 0.5   &  8&.20\p 0.21 \cr \noalign{\hrule}
  3&.78   & 20&.78\p 0.5   &  7&.33\p 0.18 \cr \noalign{\hrule}
  0&      & 19&.47\p 0.5   &  6&.49\p 0.17 \cr \noalign{\hrule}
 $-$3&.78 & 15&.90\p 0.53  &  5&.61\p 0.19 \cr \noalign{\hrule}
 $-$7&.6  & 11&.20\p 0.73  &  4&.79\p 0.31 \cr \noalign{\hrule}
$-$11&.4  &  5&.91\p 0.5   &  3&.94\p 0.33 \cr \noalign{\hrule height 0.7pt}
}}\hss}
\vskip 3pt
\caption{Results for the flow separation length.  The error is composed of the
	discretisation error of the determination of the flow reattachment
	point and a systematic error (see text).}
\label{tab:seplen}
\end{table}

The error in the separation length was calculated as follows: The determination
of the flow reattachment position for one particular simulation was accurate to
one grid spacing.  The corresponding error in the continuum limit results from
the linear regression formulas.  This error does not account for biases which
may be inherent in the turbulence model, the kind of grid and the parameter
settings used.  We estimate that systematic error in the absolute separation
length to be 0.5\,m.  The errors given in Table~\ref{tab:seplen} are the result
of adding these errors quadratically.  The systematic error dominates in most
cases.

\begin{figure}
\input{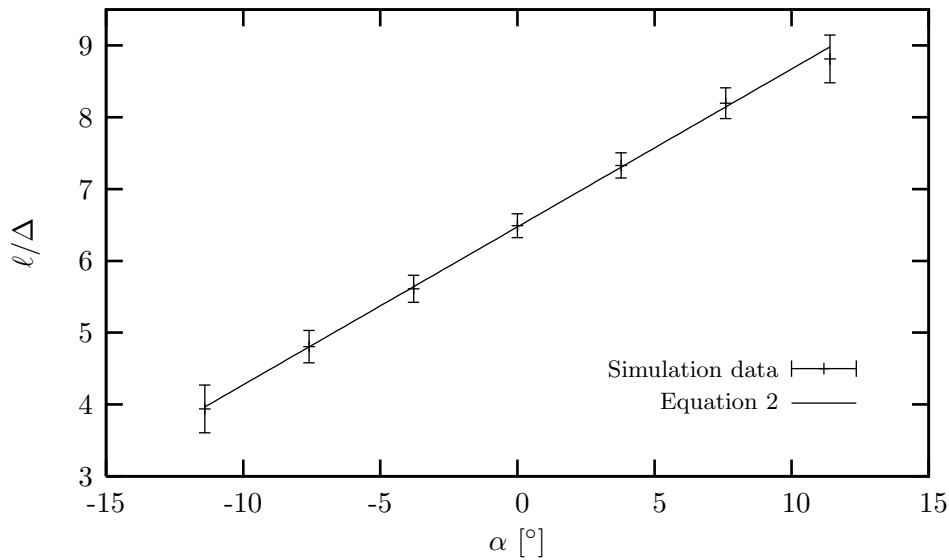}
\caption{Dependence of the non-dimensionalised separation length on the
  angle~$\alpha$.  The relationship is remarkably linear.  Note that the
  rightmost value of $\alpha$ belongs to the dune with the sharpest brink, i.\
  e.\ the shortest dune.}
\label{fig:brinkangle}
\end{figure}

Our main interest here is in the dependence of the separation length on the
dune shape.  We find that $\ell/\Delta$ is larger for dunes with a sharp brink
than for rounded dunes.  It depends linearly on the brink position~$d$,
respectively on the angle of the dune shape at the brink,~$\alpha$.  As can be
seen in Figure~\ref{fig:brinkangle}, the linear relation holds for the whole
range of brink angles investigated here.  Fitting the relation
\begin{equation}
\ell(\alpha)/\Delta(\alpha)= A\cdot \alpha + B\,,
\label{eq:brinkangle}
\end{equation}
we obtain $A=\anglefitA/^\circ$ and $B=\anglefitB$.

\begin{figure}
\input{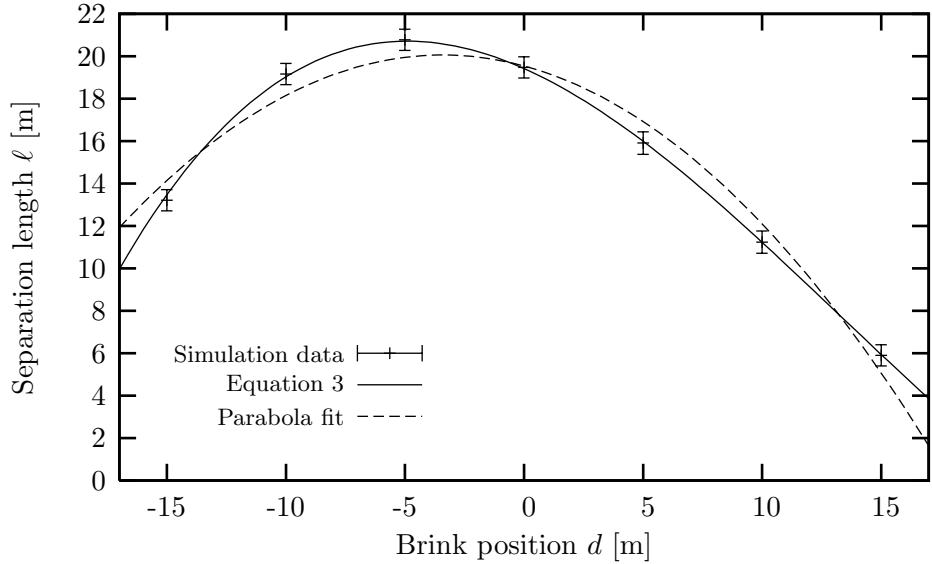}
\caption{Dependence of the absolute flow separation length on the brink
  position.  The expression derived from the linear angle dependence displayed
  in Figure~\protect{\ref{fig:brinkangle}} provides a much better fit than a
  parabola.  Note that only the dunes with $d\geq 0$ have the same height,
  while the others become smaller with decreasing $d$ (see
  Figure~\protect{\ref{fig:shape}}).}
\label{fig:brinkpos}
\end{figure}

To give the reader an idea of the actual separation lengths we obtained, we
also give our results for the absolute separation length.  The length of flow
separation decreases both for large and for very negative brink positions.  As
one can see from Figure~\ref{fig:brinkpos}, the maximum does not coincide with
$d=0$ but lies to the left of that value.  

We compute the absolute separation lengths from Equation (\ref{eq:brinkangle})
by using the geometrical relation between the brink angle $\alpha$ and the
brink position $d$.  This relation can be obtained from the geometry of our
dune profiles described above.
\begin{eqnarray}
\label{eq:anglefitpos}
\ell(\alpha(d))&=& (A\cdot\alpha(d) + B)\;\Delta(\alpha(d))
\nonumber\\
&=& \left(-A \,\arcsin \frac dR + B\right)\cdot 
\left(H_{\hbox{\scriptsize max}} 
- d\,\tan \left( \frac12 \,\arcsin\frac dR \right) \right)
\end{eqnarray}
This equation contains the crest height of the round dunes,
$H_{\hbox{\scriptsize max}}$, and the curvature radius of the dune shape at the
crest, $R$.  Both are quantities related to the set of dunes we study here, not
single dunes, and therefore stay constant during our investigation.

For curiosity we can also try a different fit to the one in
Equation~\ref{eq:anglefitpos} and compare their quality.  As the data have a
maximum and everywhere negative curvature, the most obvious candidate for a fit
is a polynomial of second order, that is a parabola.  It is plotted in
Figure~\ref{fig:brinkpos} but does not fit particularly well, even though it
has three parameters compared to two for our fit.  The angle-based fit
(\ref{eq:anglefitpos}) has a mean deviation of $\anglefitposRMS$ compared to
$\parfitRMS$ for the parabola fit.

\section{The separating streamline}
\label{sec:sepline}

In order to model the formation and evolution of sand dunes, it is necessary to
calculate the ground shear stress on which the flux of transported sand
crucially depends.  While analytic derivations of the shear stress on landforms
exist \cite{Hunt88,Weng91}, they apply to round hills from which air flow does
not separate.  The sand flux over dunes has been computed without taking into
account flow separation by Weng et al.~\cite{Weng91}.  One can go one step
further and compute the shear stress over a shape which for the most part
follows the dune shape, but coincides with the separating streamline in the
region of flow separation \cite{Sauermannphd,KroySauer03}.  Since the shape of
stationary dunes depends sensitively on the shear stress, it is of great
importance to know the shape of the separating streamline.

\begin{figure}
\hbox to \textwidth{\hss\hskip -5cm\input{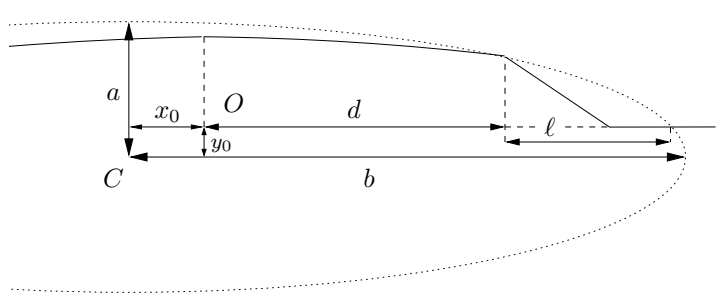}\hss}
\caption{The parametrisation of the ellipse describing the separating
streamline.  In this example, $d>0$ and both $x_0$ and $y_0<0$.  $C$ is the
centre of the ellipse, and $O$ is the origin, at ground level and at the
horizontal position of the dune crest.}
\label{fig:ellipse}
\end{figure}

From our CFD simulation, we extracted the streamline which just touches the
brink of the dune and which therefore represents a very good approximation of
the separating streamline.  In each case, we used the simulation with the
smallest grid spacing, 5\,cm.  The simulation streamline does not separate
directly at the brink, but a small distance down the slip face.  But since this
distance amounted to two grid spacings in all simulations, independent of which
grid spacing was chosen, this is a numerical effect due to the difficult
numerics at the flow separation point.  Therefore we aim to model only the part
of the separating streamline which curves downwards, not the dip near the
separation point.

\begin{figure}
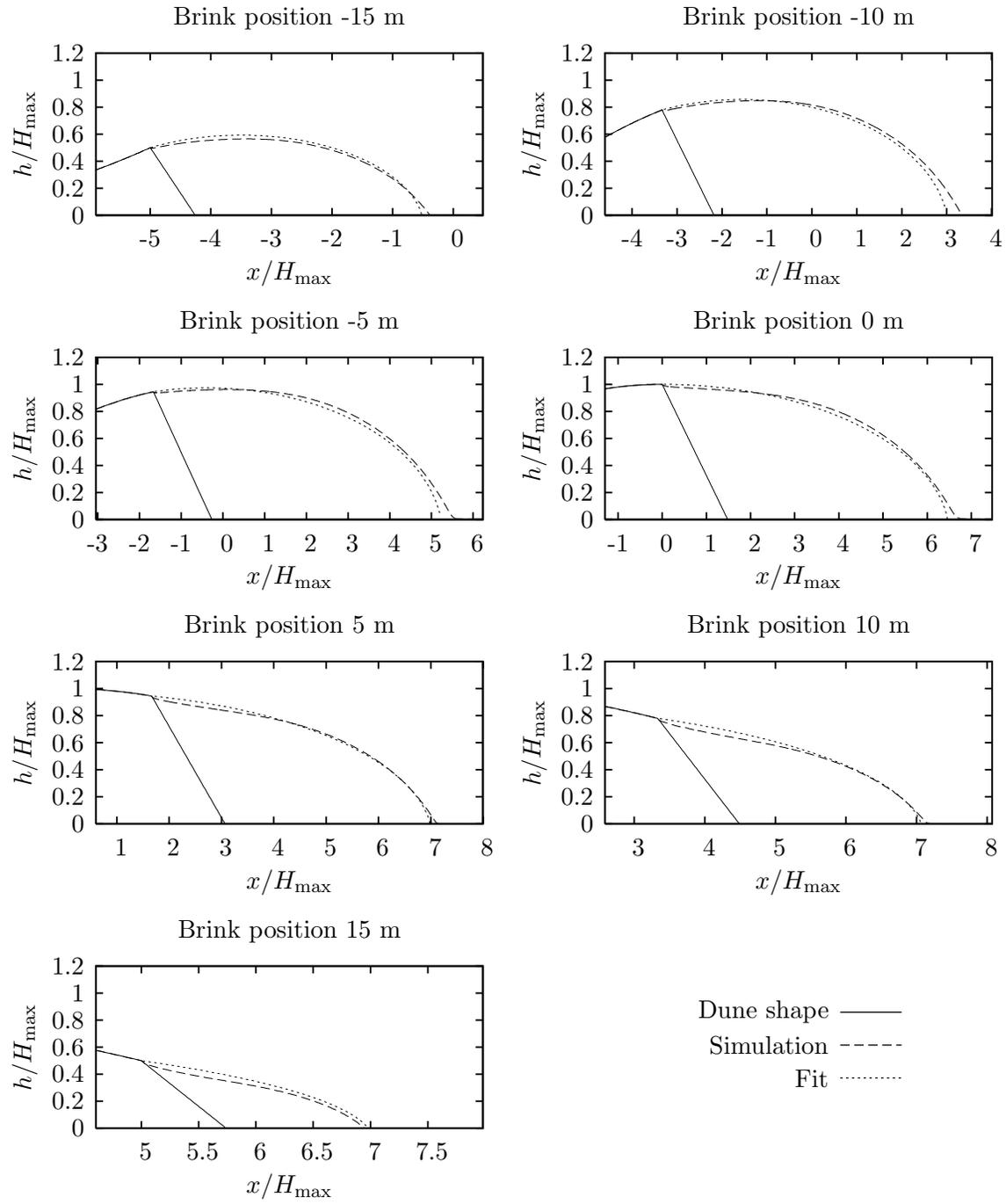

\input{allbubfit_-15.ptex}
\input{allbubfit_-10.ptex}
\linebreak
\input{allbubfit_-5.ptex}
\input{allbubfit_0.ptex}
\linebreak
\input{allbubfit_5.ptex}
\input{allbubfit_10.ptex}
\linebreak
\input{allbubfit_15.ptex}
\input{allbubfit_key.ptex}
\caption{Fit for the separating streamlines.  All coordinates are rescaled
  using the maximal height of the dunes, 3\,m.}
\label{fig:allbubfit}
\end{figure}

We found that the shape of the separating streamline is well described by an
ellipse.  An ellipse is determined by four parameters, the coordinates of the
centre and the two semiaxes (see Figure~\ref{fig:ellipse}):
\dmath2{
\frac{(y-y_0)^2}{a^2} + \frac{(x-x_0)^2}{b^2} &=& 1
&eq:ellipse}
Both the brink and the reattachment point have to lie on this ellipse, so there
remain two free parameters which have to be fitted.  We choose to fit $x_0$ and
$b$ and calculate $y_0$ and $a$ from them using the position of the brink and
the reattachment point.  The brink is located at the point $(d, \Delta)$, the
reattachment point is $(d+\ell, 0)$.  Putting these two points into the ellipse
equation (\ref{eq:ellipse}) and performing some algebra, we obtain a
biquadratic equation for $a$:
\dmath1{
a^4 + \frac{4\,\Delta^2\,b^2}{(2\,\delta+\ell)^2\,\ell^2}\,
\bigg[\,(\delta+\ell)^2-b^2 &-& \frac12\,(2\,\delta+\ell)\,\ell\,\bigg]\;a^2
+ \frac{\Delta^4\,b^4}{(2\,\delta+\ell)^2\,\ell^2} = 0\,,
&eq:a4\cr
\hbox{where}\quad\delta &=& d - x_0\,.
}
The biquadratic equation (\ref{eq:a4}) can be solved with the standard formula.
Choosing the solution for which the ellipse intersects the ground with negative
slope, we obtain:
\dmath1{
a^2 &=& 
- \frac{2\,\Delta^2\,b^2}{(2\,\delta+\ell)^2\,\ell^2}\,\bigg[\ldots\bigg]
+\sqrt{\left(\frac{2\,\Delta^2\,b^2}{(2\,\delta+\ell)^2\,\ell^2}\,
\bigg[\ldots\bigg]\right)^2 - \frac{\Delta^4\,b^4}{(2\,\delta+\ell)^2\,\ell^2}}
\;,&eq:a2}
where the expression in square brackets is the same as in Equation~\ref{eq:a4}.
Since $a$ is positive by definition, it is thereby uniquely determined.  $y_0$
can then be computed from $a$ and the constraints, giving:
\dmath1{
y_0 &=& \frac{a^2}{2\,\Delta}\,\left(\frac{\Delta^2}{a^2} - 
\frac{(2\,\delta+\ell)\,\ell}{b^2}\right)\,.
&eq:y0}

It remains to determine the unknown variables in Equation~\ref{eq:a4}.  Besides
the measures given by the geometry of the dune, the equation contains $\ell$,
$b$ and $x_0$.  $\ell$ is given by Equation~\ref{eq:brinkangle}.  The other two
quantities have to be fitted.  We obtain the best overall fit with the
following expressions:
\dmath{1}{
x_0 &=& \left\{\;\vcenter{\halign{ # \hfil & # \hfil \cr
0 &  \quad $d\geq 0$\,, \cr\noalign{\vskip -2mm}
$-7\;(H_{\hbox{\scriptsize max}}-\Delta)$ & \quad $d<0$ \cr
}}\right.
&eq:x0\cr
b &=& (d + \ell - x_0) + 0.04\cdot H_{\hbox{\scriptsize max}}
&eq:b}
It is clear that the difference between the $x$ coordinates of the ellipse's
centre and the reattachment point, $d + \ell - x_0$, is a lower bound for the
horizontal semiaxis.  The additional term in Eq.~\ref{eq:b} is required for the
ellipse to intersect the line $y=0$ at an angle rather than vertically.  It
does not depend on the brink position.

The fit together with the data from both sets of simulations is displayed in
Figure~\ref{fig:allbubfit}.  It can be seen that the fit is very accurate.  The
upward curvature of the simulation streamlines close to the brink is associated
with the delay of flow separation by two grid spacings.  This is a numerical
effect which we do not model.

\section{Closely spaced dunes}
\label{sec:close}

In the previous sections we have considered single transverse dunes.  This was
done to be able to make a statement about the shape dependence of flow
separation without at the same time dealing with complications due to potential
neighbouring dunes.  In reality, this corresponds to the case of isolated
dunes, which have a distance to their neighbours of around three times their
length or more.

\begin{figure}
\hbox to \textwidth{\hfill
\includegraphics[height=14cm,angle=-90]{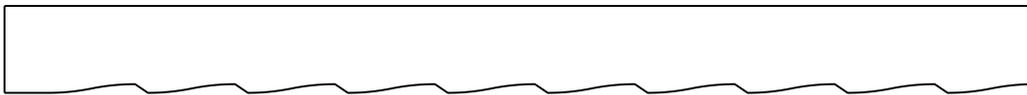}\hfill}
\caption{The simulation region in the simulation of multiple dunes.}
\label{fig:multiple}
\end{figure}

To get an idea of the influence of neighbouring dunes close by, we performed a
simulation of closely spaced dunes.  The shape of the dunes was the same as for
$d=0$ in Figure~\ref{fig:shape}.  The dunes were set next to each other so that
the foot of the upwind slope of each following dune coincided with the slip
face foot of the previous one, as shown in Figure \ref{fig:multiple}.  The
simulation parameters were the same as previously.  This simulation was only
done with the grid spacing 0.1\,m.

It should be understood that our geometrical construction leads to the upwind
side of a dune rising immediately at the foot of the previous dune.  Since the
sand cannot be moved within the separation region, this means that this profile
is not stable.  However, as we can know the length of flow separation only
after our simulation, we cannot know in advance what a stable profile would
look like.

\begin{table}
\halign{
\vrule height 2.5ex depth 1ex width 0.7pt \ # \hfil \vrule 
&\ \hfil # \hfil \vrule &\ \hfil # \hfil \vrule & \hfil\ # \hfil \vrule 
&\ \hfil # \hfil \vrule &\ \hfil # \hfil \vrule & \hfil\ # \hfil \vrule 
&\ \hfil # \hfil \vrule &\ \hfil # \hfil \vrule & \hfil\ # \hfil
\vrule width 0.7pt\cr\noalign{\hrule height 0.7pt}
Dune number & 1 & 2 & 3 & 4 & 5 & 6 & 7 & 8 & 9 \cr\noalign{\hrule}
Separation length & 17.77 & 15.78 & 15.41 & 15.19 & 15.00 & 
                    14.90 & 14.80 & 14.71 & 14.71 \cr
\noalign{\hrule height 0.7pt}
}
\caption{The separation lengths obtained from the simulation of closely spaced
        dunes.  The errors are a statistical error of 0.05\,m in the 
        determination of the reattachment point and a systematical error of 
        0.5\,m.}
\label{tab:multisep}
\end{table}

\begin{figure}
\hbox to \textwidth{\hfil\input{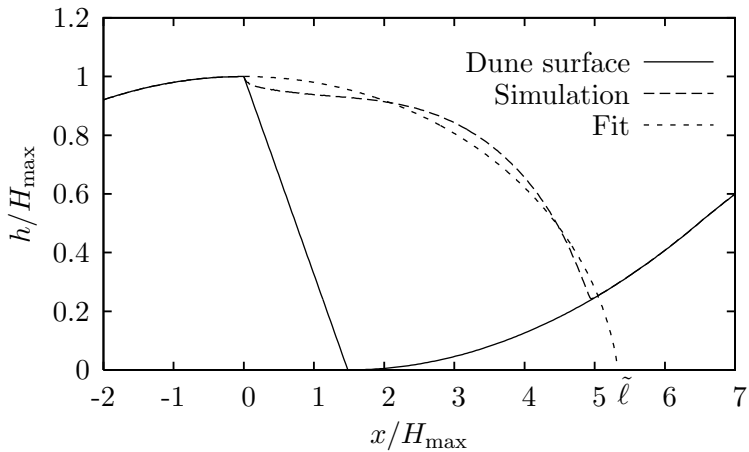}\hfil}
\caption{The ellipse fit also describes the separating streamline of closely
  spaced dunes.  This Figure shows the last of the dunes in
  Figure~\protect{\ref{fig:multiple}}.  Both $h$ and $x$ are normalised by
  dividing by the crest height, 3 m.}
\label{fig:closebubble}
\end{figure}

Table~\ref{tab:multisep} shows the separation lengths in the lee side of the
nine dunes in the simulation.  One can see that the values converge towards the
downwind end of the simulation area.  Therefore we take the separation length
of the last dune as the value for a dune in an extended dune field.  By
comparison with table~\ref{tab:seplen}, it is $\approx 25$\,\% smaller than for
an isolated dune.

We can now fit the separating streamline in the same way as for the isolated
dunes.  The Formulas (\ref{eq:ellipse}) to (\ref{eq:b}) apply unchanged except
for one modification: Since the foot of the downwind dune curves upwards, the
separating streamline intersects the ground upwind of where it would for an
isolated dune.  We account for this by replacing the separation length $\ell$
by an $\tilde\ell$, which is larger than the separation length.  $\tilde\ell$
is the intercept of the separation bubble shape with the line $h=0$, while for
closely spaced dunes the reattachment position at $x=\ell$ has a height $h>0$.
The best fit is obtained for $\tilde\ell=16\,$m.  It is shown in
Figure~\ref{fig:closebubble}.  The upward curvature of the simulated streamline
close to the brink is especially pronounced because of the large grid spacing
of this simulation.

\section{Discussion}
\label{sec:discuss}

This section compares our results to previous work.  A recent review of air
flow over transverse dunes \cite{WalkerNickling02} cites values of 4--10 for
$\ell/\Delta$.  Our results also lie within that range (see
Figure~\ref{fig:brinkangle}).  Engel \cite{Engel81} finds values for the
non-dimensionalised separation length between 4 and a little over 6, depending
on the roughness and the aspect ratio of triangular dunes.  In
Reference~\cite{Parsons04} a wide range between 3 and 15 is given for the same
quantity.  Their values are for an aspect ratio of 0.1, which applies to our
dune with $\alpha=0$, are 5.67 and 8.13, depending on the height.  This
compares well with our value of 6.49.  The discrepancy can be explained by the
different shape, in particular the fact that our dune shape for $\alpha=0$ has
a horizontal tangent at the brink, whereas the dunes in Ref.~\cite{Parsons04}
are triangular.

The field measurements \cite{Parteli04} were performed in a closely spaced dune
field.  The dune profile was measured along a straight line in wind direction,
perpendicular to the dunes.  The authors find that the distance between the
brink of each dune and the foot of the following one is typically four times
the height or below.  Under the assumption that the dune field is stationary,
this distance is an upper limit of the separation length.  We found a
separation length of 4.9 times the height for closely spaced dunes with a
horizontal tangent at the brink.  Considering that only two of the six dunes
measured in Ref.~\cite{Parteli04} had positive slope at the brink and that the
dunes with the shortest separation length were indeed very round, the agreement
is not bad.

Last but not least, our results are supported by a recent fit to experimental
data \cite{Paarlberg05}.  The authors obtain a non-dimensionalised separation
length in the range from 4 to 7.5 for a brink angle ranging from $-10^\circ$ to
$10^\circ$.  This is similar to our results, and Figure 12 in
\cite{Paarlberg05} strongly resembles our Figure~\ref{fig:brinkangle}.  The
authors present a polynomial fit for the separating streamline.  Unfortunately
a re-parametrisation of either fit which would be necessary for a quantitative
comparison is beyond the scope of this paper.  \cite{Paarlberg05} find that the
separating streamline intersects the ground at the reattachment point at an
angle.  This is contrary to what previous fits of the separating streamline
have assumed, but is again in line with our findings.

\section{Summary and outlook}
\label{sec:concl}

We have investigated the air flow over  transverse dunes of different shapes
using the commercial CFD software FLUENT.  The variation in shape was achieved
by moving the position of the slip face of the dune to different places.

We have determined the length of flow separation in the lee side of these
dunes.  For each dune shape, six simulations were performed, with two absolute
sizes of the dune and three different grid spacings to be able to remove the
remaining influences of the grid spacing.  The maximal separation length does
not occur for dune shapes with a horizontal tangent at the brink, but for
shapes with a somewhat sharper brink.  The separation length
non-dimensionalised through division by the slip face height was found to
depend linearly on the position of the slip face as represented in
Equation~\ref{eq:brinkangle}.  This linear law can be rewritten with the help
of geometric properties of the dune to give the absolute separation length.

The shape of the separating streamline, that is the boundary of the
recirculation region, is well approximated by an ellipse.  This ellipse is
constrained by the requirement that the brink and the flow reattachment point
lie on it.  The horizontal position of its centre is at the crest for rounded
dunes.  For dunes with a sharp brink, it lies to the left of the crest of the
rounded dunes, and its position is proportional to the difference in height
between the round dunes and the sharp dune in question (see
Equation~\ref{eq:x0}).  The horizontal semiaxis of the ellipse has to be chosen
so that its rightmost point lies to the right of the flow reattachment point by
0.04 times the height of the rounded dunes, independent of the brink position.

Lastly, we have extended our investigation by simulating the flow over a field
of ten closely spaced transverse dunes.  Here we restricted ourselves to one
dune shape with horizontal tangent at the brink.  The separation length reached
an asymptotic value behind the ninth dune, which was 25\,\% less than the value
for an isolated dune.

There still remain many open questions concerning the air flow over dunes.  The
most obvious restriction of our results is that they were obtained for
transverse dunes only.  The three-dimensional shape of other dunes, for
instance barchans, calls for a three-dimensional description of their
recirculation region.  Furthermore, one should investigate how accurate the
concept of a separation bubble is: It has been assumed for the purpose of sand
transport simulations \cite{KroySauer03} that the wind shear stress on a dune
shape is the same as the shear stress over a shape composed of a dune and the
recirculation region in its lee.  While good results for dune shapes support
this assumptions, it should be verified from fluid dynamics.

Lastly, the influence of the dune size should be investigated.  The flow over
dunes is fully turbulent and therefore scale invariant.  However, if the dune
is scaled up while the ground roughness and the inflow velocity profile are
kept invariant, the separation length can change.  This was found for instance
by Engel \cite{Engel81}.  It bears investigation how the phenomenological laws
and constants found in this work depend on the dune size.

\section*{Acknowledgements}

We thank Martin Winter, Jos\'e Soares de Andrade Jr.\ and Murilo Pereira de
Almeida for helpful comments and discussions and for information on the FLUENT
software.  We thank the Volkswagen Stiftung and the Max Planck Price for
funding much of our research in this field.


\end{document}